\documentclass[aps, prl, amsmath,amssymb, reprint, showpacs,superscriptaddress,groupedaddress, nofootinbib ]{revtex4-1}

\usepackage{graphicx}% Include figure files
\usepackage{dcolumn}% Align table columns on decimal point
\usepackage{bm}% bold math
\usepackage{soul}

\usepackage{color}

\usepackage{nicefrac}
\newcommand{\pb}{CH$_3$NH$_3$PbI$_3$}
\newcommand{\ba}{CH$_3$NH$_3$BaI$_3$}
\newcommand{\degc}{$^{\circ}$C} % degree celsius
\newcommand{\degree}{$^{\circ}$}

\begin{document}

\title{Crystal structure, stability and optoelectronic properties of the organic-inorganic wide bandgap perovskite \ba{}: Candidate for transparent conductor applications}
\author{Akash Kumar,$^\dagger$ K. R. Balasubramaniam}\email{bala.ramanathan@iitb.ac.in,\  aftab@iitb.ac.in}

\affiliation{Department of Energy Science and Engineering, Indian Institute of Technology Bombay, Mumbai 400076, India.}
\author{Jiban Kangsabanik,$^\dagger$ Vikram,$^\dagger$ Aftab Alam}
\affiliation{Department of Physics, Indian Institute of Technology Bombay, Mumbai 400076, India.} 
\let\thefootnote\relax\footnote{$^\dagger$ These three authors have contributed equally to this work}
\begin{abstract}
Structural stability, electronic structure and optical properties of \ba{} hybrid perovskite is examined from theory as well as experiment. Solution-processed thin films of \ba{} exhibited a high transparency in the wavelength range of 400 nm to 825 nm (1.5 eV to 3.1 eV for which the photon current density is highest in the solar spectrum)  which essentially justifies a high bandgap of 4 eV obtained by theoretical estimation. Also, the XRD patterns of the thin films match well with the \{00\emph{l}\} peaks of the simulated pattern obtained from the relaxed unit cell of \ba{}, crystallizing in the I4/mcm space group, with lattice parameters, \emph{a} = 9.30 \r{A}, \emph{c} = 13.94 \r{A}. Atom projected density of state and band structure calculations reveal the conduction and valence band edges to be comprised primarily of Barium \emph{d}--orbitals and Iodine \emph{p}--orbitals, respectively. The larger band gap of \ba{} compared to \pb{} can be attributed to the lower electro-negativity coupled with the lack of \emph{d}--orbitals in the valence band of Ba$^{2+}$. A more detailed analysis reveals the excellent chemical and mechanical stability of \ba{} against humidity, unlike its lead halide counterpart, which degrades under such conditions. We propose La to be a suitable dopant to make this compound a promising candidate for transparent conductor applications, especially for all perovskite solar cells. This claim is supported by our calculated results on charge concentration, effective mass and vacancy formation energies.
\end{abstract}


\maketitle

Recently, compounds in the organic-inorganic halide perovskite family (\emph{AB}X$_3$: \emph{A} is an organic cation, \emph{B} is an inorganic cation, and X is a halide element) have garnered a lot of attention in the solar photovoltaic community. This is due to their superior optoelectronic properties,  easy synthesis techniques and variety of compounds that can be obtained \emph{via} simple substitutions of the \emph{A}, \emph{B} and X ions. Specifically, solar cells, with (CH$_3$NH$_3$)$^+$ as the \emph{A} cation and Pb$^{2+}$ as the \emph{B} cation, have shown a rapid growth in the solar-to-electricity power conversion efficiency.\cite{burschka2013sequential, lee2012efficient, zhou2014interface, kieslich2014solid, wehrenfennig2014charge}  The lead halide perovskite solar cell was first introduced by Kojima \emph{et al} in 2009, wherein it was used in a dye-sensitized solar cell architecture.\cite{kojima2009organometal} Much of the research in recent times has focused on solid-state cells with different architectures, hole transport layers, compositional engineering, and synthesis techniques. \cite{burschka2013sequential, zhou2014interface, kim2012lead, jeon2014solvent, jeng2013ch3nh3pbi3, wang2014large} Even then, there are some caveats associated with the various components of the CH$_3$NH$_3$PbX$_3$-based solar cells: the stability of the absorber material in ambient conditions and the presence of Pb to name a couple. Active research to address these problems is being conducted worldwide through suitable replacements to both the CH$_3$NH$_3$ and Pb cations.

Tunability of the properties by changing the constituent elements gives this class of material more scope of research and applicability.\cite{edri2013high, nie2015high, wehrenfennig2014homogeneous} Such tunability in the bandgap has been observed in the oxide perovskites, wherein only a small variation in bandgap is observed in \emph{A}TiO$_3$ compounds, with A being Ca (3.46 eV), Sr (3.25 eV), Ba (3.2 eV) or Pb (3.4 eV).\cite{balachandran1982electrical, koch1975bulk, piskunov2004bulk} However, there is a large change in bandgap on changing the B cation, e.g. BaTiO$_3$ (3.2 eV) and BaZrO$_3$ (5.3 eV). \cite{robertson2000band} The optimization of the absorber material in the case of the lead halide perovskites is also done in a similar manner, albeit \emph{via} replacement of the halide \emph{i.e.} I with Br and Cl and the associated changes in the band gap and structure was the subject of many studies.\cite{jeon2015compositional, noh2013chemical, edri2014chloride, schmidt2014nontemplate}  Similarly, reports on replacing or mixing organic A cation i.e, CH$_3$NH$_3^+$ with formamidinium have shown that their stability in ambient conditions is increased, albeit its impact on properties such as carrier mobility and band gap is not that significant ($\approx$ 0.1 eV).\cite{eperon2014formamidinium, yang2016effects, jeon2015compositional}

In this letter, we will discuss the effects of complete replacement of Pb by Ba in \pb{}. When Shannon ionic radius of cation A(CH$_3$NH$_3$) and anion X(I) are taken as 1.8 \r{A} and 2.2 \r{A} respectively,\cite{shannon1969effective} a stable high symmetric tetragonal structure (energetically most favourable for \pb{}) requires the radius of cation B to be in-between 0.94 and 1.84 \r{A} as per Goldschmidt's tolerance factor estimate. Charge neutrality and coordination number of 6 in the BX$_6$ octahedra further reduces the number of possible replacements at B site\cite{yin2015halide, mitzi2001structurally} Considering the size to be comparable to Pb, Ba and Sr are obvious choices for the Pb replacement at the B site.  Sr as a replacement for Pb has already been reported.\cite{jacobsson2015goldschmidt} Using the ionic radii of MA cation (A) , Ba, and I as 1.80 \r{A}, 1.35 \r{A} and 2.2 \r{A} respectively,\cite{jacobsson2015goldschmidt} the Goldschimdt tolerance factor is calculated to be 0.797 for the \ba{}  crystallizing in the perovskite structure. We, therefore, expect the crystal structure and the atom positions of \ba{} to be similar to that of the tetragonally distorted perovskite \pb{} (Goldschmidt tolerance factor = 0.83).\cite{brivio2015lattice} Here, organic-inorganic barium halide perovskite was synthesized \emph{via} solution processing and its XRD and UV-Vis spectroscopy was done. The ab-initio density functional calculation validates our experimental data as well as explores other properties such as chemical and mechanical stability, electronic structure and advantages of La doping.

\begin{figure}[t]
\includegraphics[width=0.45\textwidth]{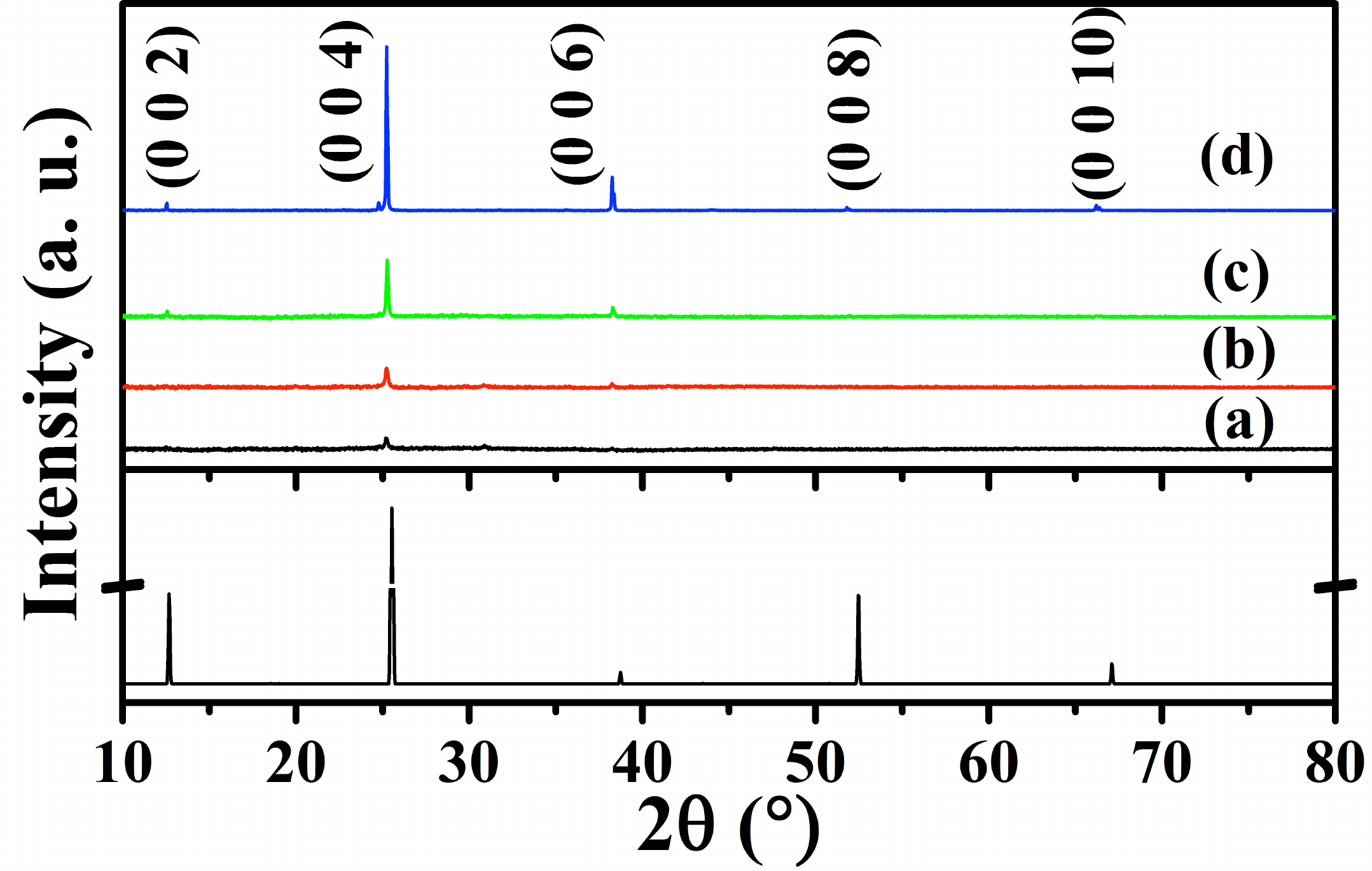}
\caption{(Color online) XRD patterns of the \ba{} films obtained on glass substrates from solution held under ambient conditions for (a) 7 days, (b) 14 days, (c) 21 days, and (d) 25 days. The bottom pattern is simulated XRD pattern from our \emph{ab initio} calculated structure of \ba.}
\label{fig:xrd}
\end{figure}

Equimolar amounts of the synthesized CH$_3$NH$_3$I powder (see \cite{suppl}) and anhydrous BaI$_2$ (99.995\% trace metal-Sigma Aldrich) were dissolved in Di-methyl formamide (DMF) to form a solution with 40 wt\% of \ba{}. The freshly prepared solution was maintained at a temperature of 70 \degree{}C under constant stirring in a glove box. Part of the solution was withdrawn from this solution repository after different time periods: 12 h, 24 h, 48 h, and 96 h. Thin films were prepared \emph{via} spin coating from these solutions followed by an annealing step at 100 \degc{}, before XRD analysis was done on these samples. Unlike \pb{}, there were no XRD peaks observed, under any of the conditions. The XRD patterns of these thin films prepared with the solution withdrawn after 12 h, 24 h, 48 h, and 96 h are shown in Fig. S1.1 of supplement.\cite{suppl} Our observations are similar to that of the synthesis of CH$_3$NH$_3$SrI$_3$.\cite{jacobsson2015goldschmidt} The difficulty in the synthesis of the perovskite, where the B cation is an alkaline earth element, can be attributed to the orthorhombic structure of SrI$_2$ inhibiting the incorporation of CH$_3$NH$_3$I and further intercalation of layers. In the case of \pb{}, the rhombohedral PbI$_2$ cell easily traps a CH$_3$NH$_3$I molecule in its voids.

A freshly prepared 40 wt\% solution was kept at 70 \degree{}C under constant stirring for 12 h, in the glove box. The solution was taken out afterwards and kept under the ambient conditions in a closed vial. This solution was then spin-coated and annealed at 100 \degree{}C after regular intervals of 7, 14, 21, and 25 days. The XRD pattern observed is shown in Fig.~\ref{fig:xrd}. The XRD patterns of the films made on 7$^{\mathrm{th}}$ (Fig.~\ref{fig:xrd}(a)) 14$^{\mathrm{th}}$ (Fig.~\ref{fig:xrd}(b)), 21$^{\mathrm{st}}$ day (Fig.~\ref{fig:xrd}(c)), and 25$^{\mathrm{th}}$ day (Fig.~\ref{fig:xrd}(d)) have distinct XRD peaks, indicating that ample time is required for the crystallization of the compound. Crystallization of the film is first observed after 7 days where in a peak at 2$\theta$ = 25.29\degree{} is seen. Longer wait times resulted in  XRD pattern with increased peak intensity at 2$\theta$ = 25.29\degree{}. In addition, we also observe appearance of other peaks at 2$\theta$ = 12.60\degree{}, 38.31\degree{}, 51.86\degree{}, and 66.26\degree{}. The corresponding \emph{d}-spacing values, calculated from the 2$\theta$ positions of the five peaks (in Fig.~\ref{fig:xrd}(d)) are 7.01 \r{A}, 3.53  \r{A}, 2.34  \r{A}, 1.76 \r{A}, and 1.40 \r{A}. One suggestive possibility from this analysis is that the \emph{d}-spacings correspond to a single $\{hkl\}$ family of planes; the spin coating resulted in a textured film. \ba{} being a novel compound, does not have a reference JCPDS file so we resorted to first principle calculations to determine the most stable structure for \ba{}.

\begin{figure}[t]
\includegraphics[width=0.33\textwidth]{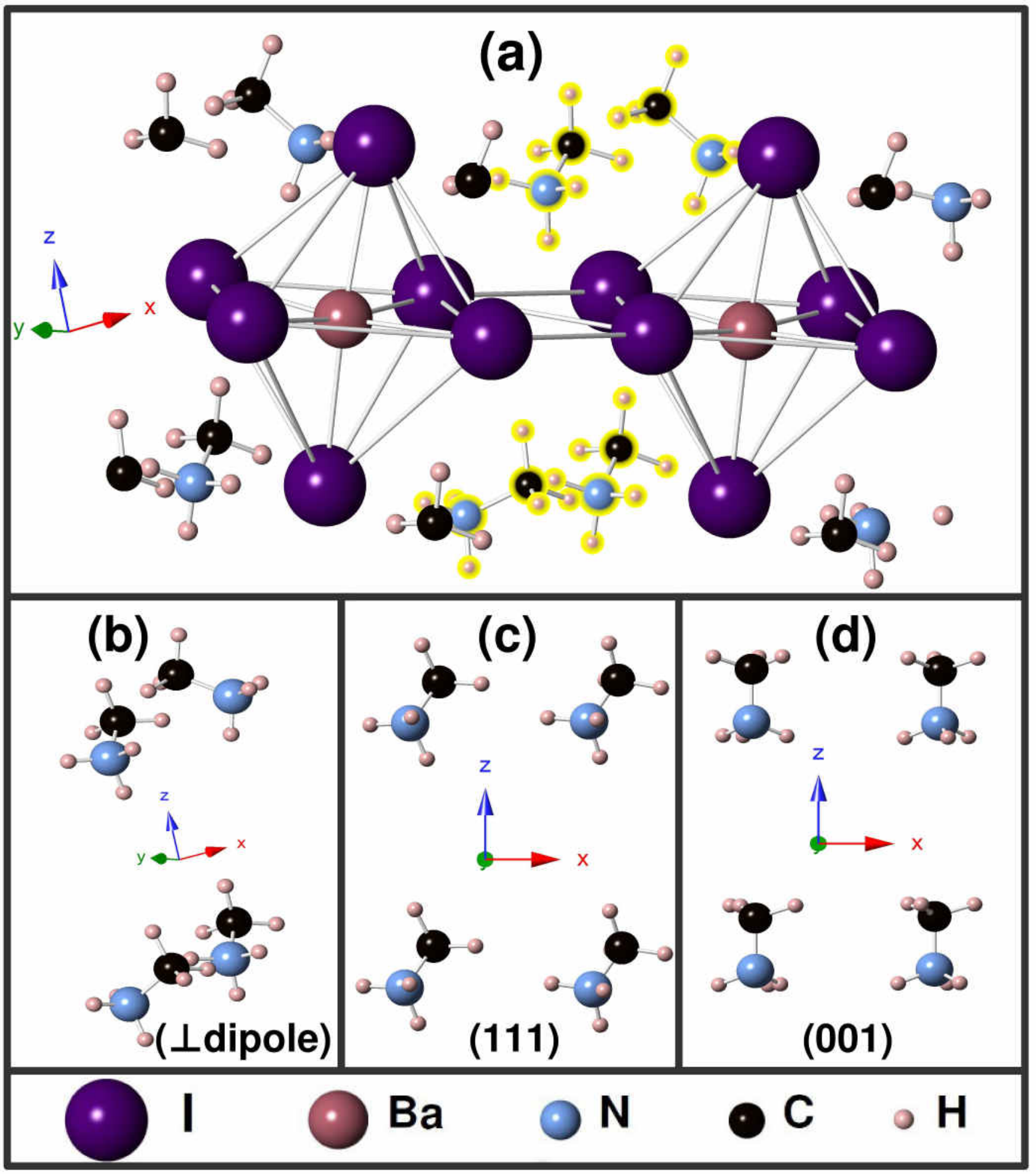}
\caption{ (Color online) (a) Relaxed structure of \ba{}. The C-N dipoles in the organic molecule, as highlighted in (a), are relaxed in three different orientations i.e when they are  aligned (b) $\perp$ (c)  along (111) and (d) (001) directions.}
\label{fig:struct}
\end{figure}

First principle calculations were performed using Density Functional Theory (DFT) with plane wave basis set as implemented in Vienna Ab initio Simulation Package (VASP).\cite{kohn1965self, kresse1996efficiency, kresse1999ultrasoft} Projector Augmented Wave (PAW) pseudo-potentials with Perdew--Burke--Ernzerhof (PBE) exchange correlation functional were used.\cite{blochl1994projector,perdew1996generalized} A plane wave energy cut off of 500 eV with a gamma centered k-mesh was used to sample the Brillouin zone. Forces (energies) were converged to values less than 0.01 eV/\r{A} ($10^{-6}$) eV). Further computational details are provided in the Supplement.\cite{suppl}

Figure~\ref{fig:struct}(a) shows the representative crystal structure of \ba{} (space group I4/mcm ($\# 140$)) with highlighted organic molecules.  As the orientation of organic molecule CH$_3$NH$_3^+$ plays a key role in stabilizing these class of compounds,\cite{motta2015revealing} we fully relaxed the structure with three different orientations of C-N dipoles, as shown in Fig.~\ref{fig:struct}(b), (c), and (d). These are basically when neighboring C-N dipoles are (b) $\perp$ to each other (c) along (111) and (d) (001) directions respectively. The (001) and (111) orientation of C-N slightly change their alignment after relaxation. The structure having $\perp$ neighboring C-N dipoles turns out to be the most stable structure with (001) and (111) oriented structures having energies 4.8 meV/atom and 6.3 meV/atom higher than this, respectively. This was also the case for \pb.\cite{brivio2015lattice} The relaxed lattice parameters for most stable structure came out to be a= 9.299 \r{A}, b= 9.301  \r{A} and c= 13.936  \r{A} which are larger than that of \pb{} as expected because  of a larger ionic radius of Ba than Pb.\cite{shannon1969effective} Notably, the structure deviates slightly from perfect tetragonal shape.

Based on the atom positions, lattice parameters, C--N bond alignment, and the space group, the XRD pattern for \ba{} with a \{00\emph{l}\}-texture was simulated with the CrystalMaker software suite [CrystalMaker Software Ltd, Begbroke, Oxfordshire, England]. Such a pattern is shown in bottom panel of Fig.~\ref{fig:xrd}. For the simulated pattern, the \{00\emph{l}\} peaks are found at 2$\theta$ = 12.69\degree{}, 25.5\degree, 38.74\degree{}, 52.49\degree{}, and 67.11\degree{}, which are in good agreement with the 2$\theta$ positions of the peaks in the XRD pattern for the film obtained after the 25 days [Fig.~\ref{fig:xrd}(d)]. The \emph{c} lattice parameter calculated from the \emph{d}--spacings of the various \{00\emph{l}\} peaks for the simulated $[\emph{c} =  13.94 \, \mathrm{\r{A}}]$ and the experimental $[\emph{c} = 14.08 \pm 0.03 \, \mathrm{\r{A}}]$ pattern differ by $\approx$ 1\%, which can possibly be attributed to a tensile strain along the growth direction.

\begin{table}[b]
\caption{Elastic constants of tetragonal \ba (GPa) }
\centering
\begin{tabular}{cccccc}\hline
  C$_{11}$\hspace{0.8cm} & C$_{33}$\hspace{0.8cm}  & C$_{44}$\hspace{0.8cm}  & C$_{66}$\hspace{0.8cm} & C$_{12}$\hspace{0.8cm}  &  C$_{13}$ \\ \hline
$12.58$\hspace{0.7cm}  & $21.56$\hspace{0.7cm} & $6.29$\hspace{0.7cm} & $~3.18$\hspace{0.7cm} & $9.52$\hspace{0.7cm} & $6.11$ 
\\ \hline
\end{tabular}
\label{table1}
\end{table}

\begin{figure}[t]
\includegraphics[width=0.5\textwidth]{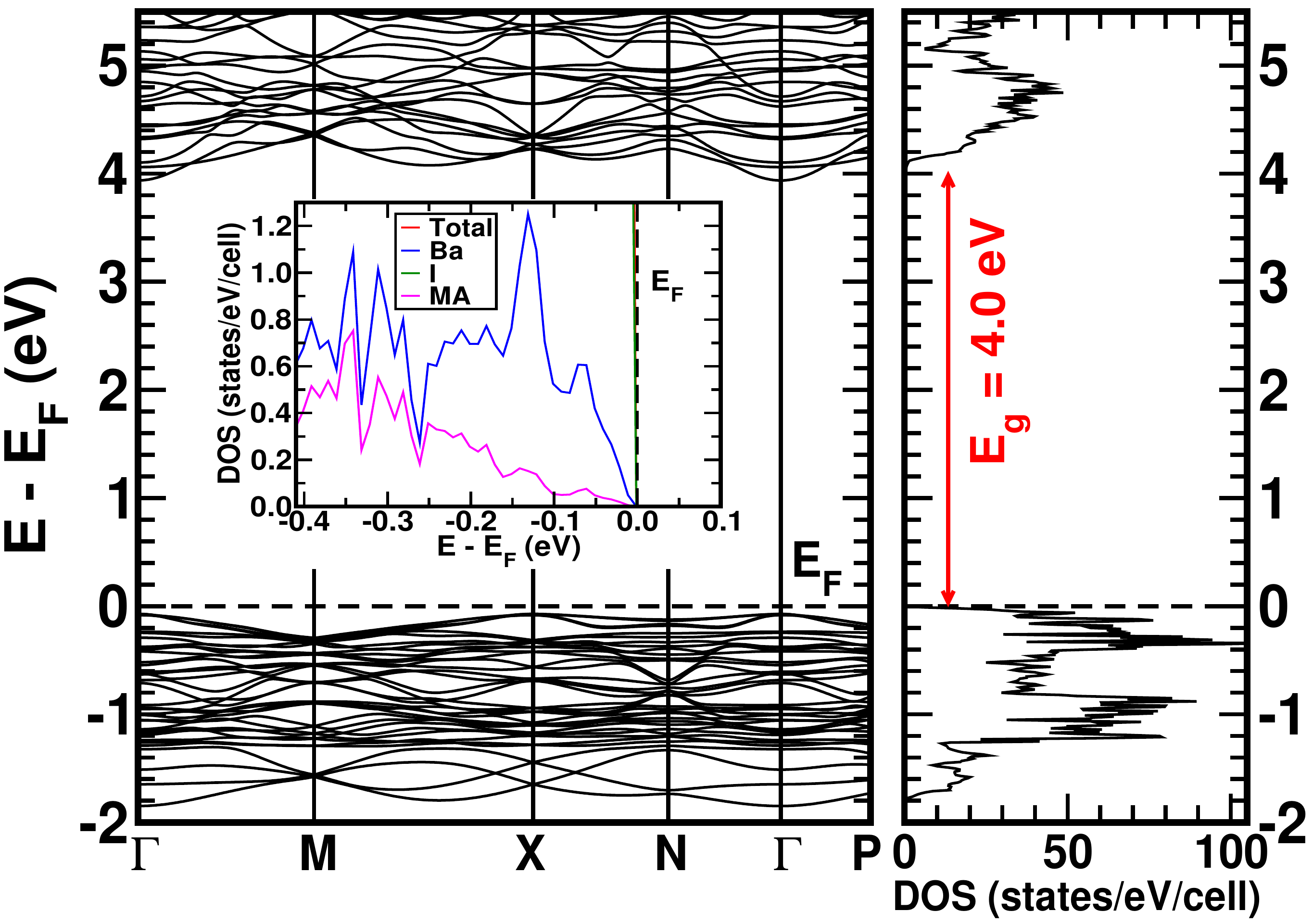}
\caption{(Color online) Band structure and total density of states (DOS) for \ba{} at  relaxed lattice parameters. The valence band near E$_F$ are majorly contributed by Iodine p-orbitals (see inset for atom projected DOS) while those near conduction bands are composed of Ba d-orbitals.}
\label{fig:band}
\end{figure}

For comparison sake, we calculated the formation energies of CH$_3$NH$_3B$X$_3$ for \emph{B} = Ba, Pb, Ca and Sr (see Section 4 of Supplement\cite{suppl}). Ba based compound turn out to be quite stable. Apart from chemical stability, there are always concerns regarding the mechanical stability of these class of compounds. In order to check that, we have done a lattice dynamics calculation to estimate the elastic properties of \ba{} (see \cite{suppl} for computational details). Table~\ref{table1} shows the elastic constants for tetragonal \ba. These elastic constants satisfy the Born and Huang mechanical stability criteria\cite{born1954dynamical} for a tetragonal structure, i.e. $C_{11} > 0$, $C_{33} > 0$, $C_{44} > 0$, $C_{66} > 0$, $(C_{11} - C_{12}) > 0$, $(C_{11} + C_{33} - 2 \times C_{13} ) > 0$, and $[2 \times (C_{11}+ C_{12}) + C_{33} + 4 \times C_{13}] > 0$, implying the strong mechanical stability of the structure. The Voigt and the Reuss bound of the bulk modulus and shear modulus are 10.16, 9.85, 4.82, 3.48, respectively in GPa.\cite{wu2007crystal}

Interestingly, XRD analysis for the same film kept for 20 days in normal humid conditions gave the previously obtained pattern. This was something unusual compared to \pb{}, which degrades in humid atmosphere, suggesting \ba{} to be stable even in humid environment.

Figure ~\ref{fig:band} shows band structure and density of states (DOS) for \ba{}. It shows a band gap of 4.0 eV, which is in fair agreement with  CH$_3$NH$_3B$I$_3$ (B=Alkaline earth) class of materials. \cite{jacobsson2015goldschmidt} Analysis of the band topology and atom projected DOS (inset) near the Fermi level (E$_F$) reveals that the valance band edges have main contribution from the Iodine \emph{p}--orbitals whereas the Ba d-orbitals contributes majorly to the conduction band edges (see Fig. S5.1 of Supplement.\cite{suppl}). The larger band gap in this case as compared to \pb{} can be mainly attributed to the much lower electronegativity of Ba than Pb.\cite{grote2014tuning, wang2014band}
\begin{figure}[t]
\includegraphics[width=0.48\textwidth]{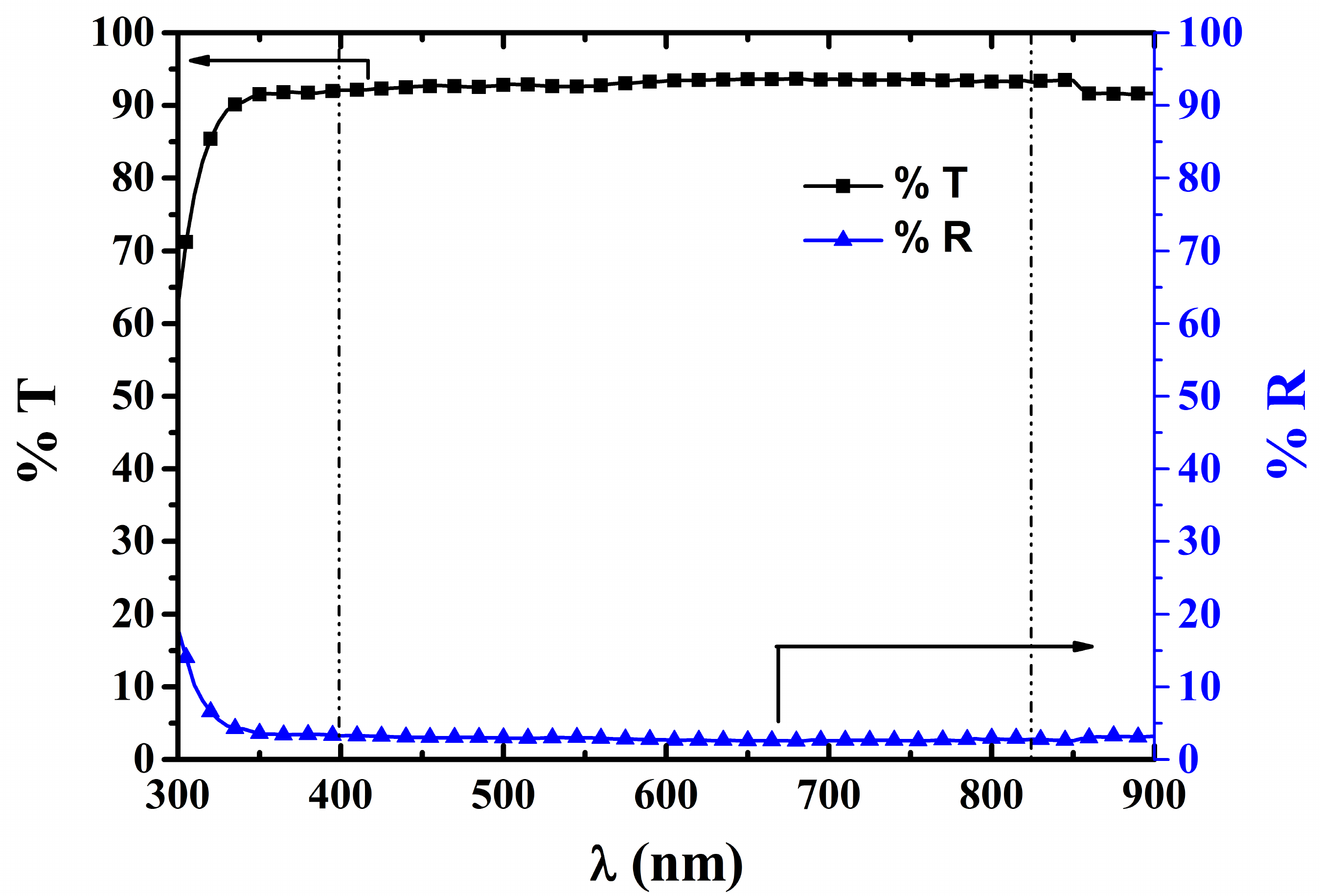}
\caption{(Color online)Transmission and reflectance spectra of films obtained on glass substrates \emph{via} spin coating of the precursor solution after waiting for 25 days. A transmittance of 90 \% and more is observed in the demarcated wavelength range of 400 nm to 825 nm for the film.}
\label{fig:trans}
\end{figure}

The high bandgap of \ba{} indicated by our \emph{ab initio} calculations implies that this material should show high transmittance in the visible range, thus suggesting its use as the base material for developing a perovskite transparent conductor (TC). Towards this, the precursor solution that was held for 25 days, was spin-coated on soda lime glass and the transmission and reflection were measured using UV-Vis spectrometer. Both the reflection and transmission spectra were recorded after applying the baseline correction of bare glass substrates. It is evident from Fig.~\ref{fig:trans} that the film's transmission in the wavelength range of 400--825 nm is always higher than 90\% and reflectance is always less than 5\%. Efforts are ongoing in our lab to develop experimental techniques to achieve doping of this highly transparent, insulating material to make it conducting.

\begin{figure}[t]
\includegraphics[width=0.49\textwidth]{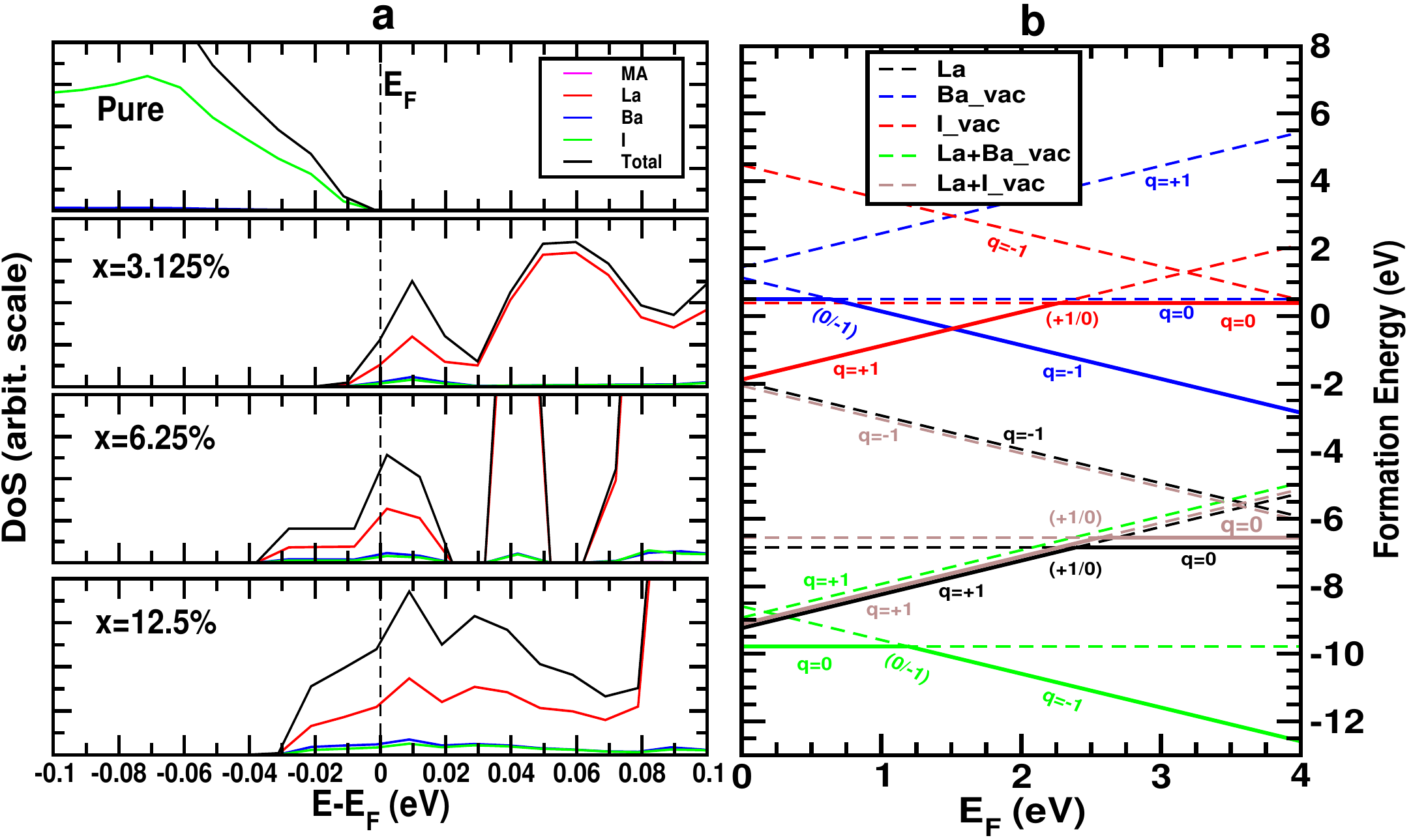}
\caption{(Color online) (a) Total and atom projected  density of states for CH$_3$NH$_3$Ba$_{1-\text{x}}$La$_{\text{x}}$I$_3$ (x$=0, 3.125, 6.25$ and $12.5 \%$) at relaxed lattice parameters. (b) Vacancy formation energy vs. E$_F$ for single vacancy defect at Ba and I sites in parent and $3.125\%$ La doped compound. Black lines indicate the result for doped compound without any vacancy. Each defect is simulated in three charged states (q=$-1, 0, +1$)}
\label{fig:dope}
\end{figure}

Historically, oxide perovskites (band gap in UV range)  are often used as transparent conductors with optimal doping. There has been some efforts to utilize hybrid perovskites for the same purpose.\cite{jacobsson2015goldschmidt} To explore this possibility, we simulated the effect of La doping i.e. CH$_3$NH$_3$Ba$_{1-x}$La$_{x}$I$_3$,which is achieved by replacing $1$, $2$ and $4$ Ba atoms with La in a $2\times2\times2$ supercell which gives $3.125\%$, $6.25\%$ and $12.5\%$ La-doping respectively. Figure~\ref{fig:dope}(a) shows the total and atom projected DOS for pure and La-doped compounds. It should be noted that, La doping introduces small states at E$_F$ and more importantly shifts it towards the conduction band. The separation between Iodine \emph{p}--orbitals in the valence band and Ba d-orbitals in the conduction band almost remains same suggesting the formation of a donor level close to the conduction band keeping the band gap almost same, and hence reflecting a degenerate semiconductor. Naively, one can imagine a figure of merit to be ($\sigma/\alpha$) for designing a TC, where $\sigma$, $\alpha$ are the electrical conductivity and absorption coefficient respectively. Empirically, $\sigma$ is related to charge concentration ($n_c$) and effective mass ($m^{*}$) as, $\sigma =n_{c} e^2 \tau /m^*$. As such, within a non-varying relaxation time approximation, the ratio ($n_c / m^*$) plays an important role in the TC design. These quantities along with the band gap as a function of various La-doping are shown in Fig. S6.1 of supplement.\cite{suppl} Notably, the ratio ($n_c / m^*$) increases with increasing doping concentration ($x$) and hence the conductivity.

Doping, however, enhances the possibility of defect formation and often reduce the mobility. Figure \ref{fig:dope}(b) shows the vacancy formation energies (see Section 4 of Supplement\cite{suppl}) as a function of E$_F$ for various single vacancy defects (i.e. Ba and I vacancies in parent and $3.125\%$ La-doped compounds). Various colored lines indicate different defects. There are $3$ charged states (q=-1,0,+1) for each defects. Solid lines indicate defect charge state at minimum energy. Points of transition of charge states is indicated as (q$_1$/q$_2$), e.g. (0/-1) means charge transition occurs from state 0 to -1. Notably, Ba vacancy in La-doped sample (green line) is the lowest in energy, and hence are most likely to form. Vacancy free La-doped compound and the same compound with Iodine vacancy are almost equally favorable. Ba and I vacancy in the parent compound are least favorable, as expected. Vacancies in La doped sample can thus play an important role in conduction. 

In summary, \ba{} was synthesized by solution processing. Relatively longer time ($\geq$ 7 days) was required for obtaining a proper mixture in the precursor solution leading to a crystalline film upon spin coating. One possible reason, suggested by the simulation data, is that the formation energy needed to crystallize the compound is higher compared to the \pb{}. The high transmittance in visible range, a remarkable stability and our calculated results based on charge concentrations, effective masses and vacancy formation energies for La-doped \ba{} opens up the possibility of using this material in applications such as transparent conductors, scintillator detectors \emph{etc}.. 
\begin{acknowledgments}
This paper is based on work supported under the US-India Partnership to Advance Clean Energy-Research (PACE-R) for the Solar Energy Research Institute for India and the United States (SERIIUS), funded jointly by the U.S. Department of Energy (Office of Science,Office of Basic Energy Sciences, and Energy Efficiency and Renewable Energy, Solar Energy Technology Program, under Subcontract DE-AC36-08GO28308 to the National Renewable Energy Laboratory, Golden, Colorado) and the Government of India, through the Department of Science and Technology under Subcontract IUSSTF/JCERDC-SERIIUS/2012 dated 22$^{\mathrm{nd}}$ Nov. 2012.
\end{acknowledgments}

\end{document}